\documentclass[a4paper]{article}
\usepackage{float}
\usepackage{multirow}
\usepackage{algorithm}
\usepackage{algpseudocode}
\usepackage{amsmath}
\usepackage[top=2cm, bottom=2cm, left=2cm, right=2cm]{geometry}
\usepackage{algorithmicx}
\usepackage{algpseudocode}

\usepackage{INTERSPEECH2022}

\title{Interactive Audio-text Representation for Automated Audio Captioning with Contrastive Learning}
\name{Chen Chen$^1$, Nana Hou$^{1\ast}$\thanks{$\ast$Nana Hou contributed to this work before leaving Nanyang Technological University, Singapore.}, Yuchen Hu$^1$, Heqing Zou$^1$, Xiaofeng Qi, Eng Siong Chng$^1$}
\address{
  $^1$School of Computer Science and Engineering, Nanyang Technological University, Singapore
 }
\email{chen1436@e.ntu.edu.sg}

\begin{document}

\maketitle
\begin{abstract}
Automated Audio captioning (AAC) is a cross-modal task that generates natural language to describe the content of input audio. Most prior works usually extract single-modality acoustic features and are therefore sub-optimal for the cross-modal decoding task. In this work, we propose a novel AAC system called CLIP-AAC to learn interactive cross-modality representation with both acoustic and textual information. Specifically, the proposed CLIP-AAC introduces an audio-head and a text-head in the pre-trained encoder to extract audio-text information. Furthermore, we also apply contrastive learning to narrow the domain difference by learning the correspondence between the audio signal and its paired captions. Experimental results show that the proposed CLIP-AAC approach surpasses the best baseline by a significant margin on the Clotho dataset in terms of NLP evaluation metrics. The ablation study indicates that both the pre-trained model and contrastive learning contribute to the performance gain of the AAC model.

\end{abstract}
\noindent\textbf{Index Terms}: Automated audio captioning, cross-modal translation, pre-trained model

\section{Introduction}
Automated Audio Captioning (AAC) aims to generate descriptive sentences for audio inputs as a cross-modal translation task. It serves as a core module in many real-world applications, such as producing text labels for sound search engines~\cite{liu2021cl4ac} and assisting the hearing impaired to access audio content~\cite{mei2021encoder}.

Most AAC approaches typically adopt encoder-decoder structures~\cite{drossos2017automated}, where the audio encoder extracts acoustic features from raw audio inputs, and the text decoder generates corresponding descriptive captions. Recent work~\cite{xu2021investigating} observed that it is difficult to train a strong encoder for audio inputs because the supervision only comes from captions, which is quite limited. To overcome such problem, prior studies~\cite{mei2021encoder, xu2021sjtu, wu2020audio, chen2020audio, mei2021audio, kong2020panns} proposed transfer learning to pre-train audio encoders on Audioset~\cite{gemmeke2017audio} for better acoustic features. However, such acoustic features only contain audio information, which has no text information as a bridge for decoders and thus leads to misalignment of audio-text information~\cite{koh2021automated}. 

In the computer vision field, recent work~\cite{radford2021learning} proposed a novel image-text pre-trained model, known as CLIP, which was trained with millions of image-text pairs and can therefore offer better features rich in semantics and visuals for cross-modal task~\cite{mokady2021clipcap}. The prior study is the source of inspiration for this work.

In this paper, we propose a novel encoder-decoder method, named CLIP-AAC, to learn the cross-modality embeddings with both acoustic and textual information for the AAC task. Specifically, the encoder in CLIP-AAC consists of an audio-head and a text-head, which are designed to learn the audio embeddings and text embeddings from audio inputs and corresponding captions. Furthermore, we propose contrastive learning to enhance the correspondence between audio-text embeddings. As a result, the extracted features can learn both audio and textual information, which are then fed into the decoder to generate descriptive sentences. At the training stage, the caption provides supervision for both features extraction and token prediction. At the inference stage, as the caption input is not available, the text-head of the encoder is discarded. Experimental results show that the proposed CLIP-AAC approach surpasses strong baselines on the Clotho dataset in terms of NLP evaluation metrics. The ablation study indicates that both the pre-trained model and contrastive learning contribute to the performance gain of the CLIP-AAC model.

The rest of this paper is organized as follows. In section 2, we introduce the proposed CLIP-AAC architecture. In section 3, experimental settings and evaluation metrics are presented. Section 4 reports and analyzes the result of the experiments. Section 5 concludes the study.\par
 
\section{CLIP-AAC Architecture}
\begin{figure*}[t]
\begin{center}
\includegraphics[scale=0.7]{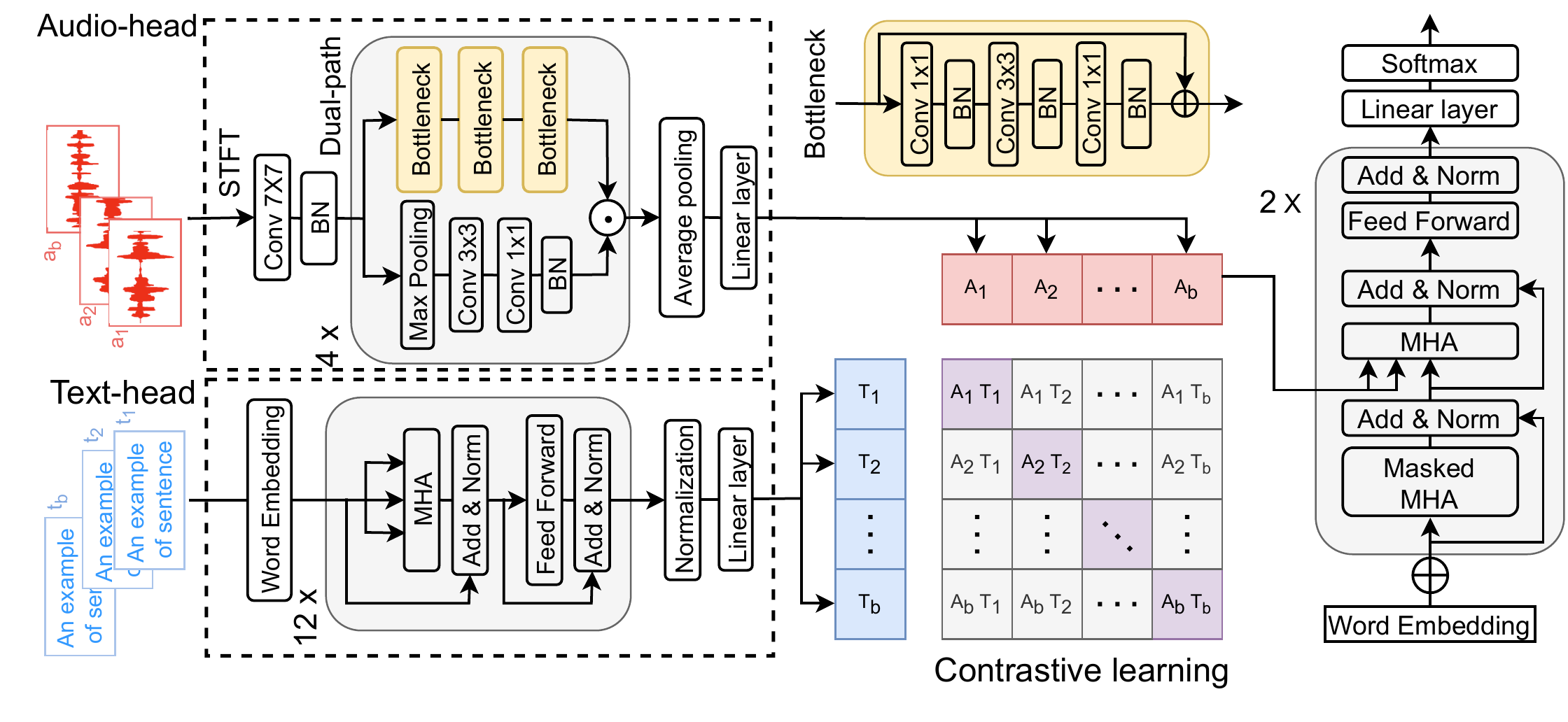}
\end{center}
\vspace{-0.16in}
\caption{The block diagram of the proposed CLIP-AAC model. The two dashed boxes denote the audio-head and text-head, respectively. ``BN" is the batch normalization layer, and ``MHA" denotes the multi-head attention mechanism. \{$A_1$,$A_2$,...,$A_b$\} and \{$T_1$,$T_2$,...,$T_b$\} are the audio embeddings and text embeddings extracted from audio-text pairs.}
\vspace{-0.1in}
\label{CLIP-AAC}
\end{figure*}

We now introduce the proposed CLIP-AAC architecture, which
consists of three modules: the encoder, the contrastive learning module and the decoder, as illustrated in Figure~\ref{CLIP-AAC}.

\subsection{Audio-head and Text-head of Encoder}
Given the $N$ audio-text pairs, we denote the AAC dataset as $D=\{(a_i,t_i)\}, i \in \{1, 2, ..., N\}$, where $a$ is the time-domain audio clip and $t$ is the caption to describe this audio. In a typical AAC model, the encoder $Enc$ is designed to convert the audio signal to the audio embedding $A = Enc(a)$. Such approach usually extracts single-modality acoustic features and are therefore sub-optimal for the cross-modal decoding task.

To alleviate such an issue, we propose an audio-head $Audio\_Enc$ and a text-head $Text\_Enc$ for the encoder in CLIP-AAC as shown in Figure~\ref{CLIP-AAC}, which take both audios and captions as inputs to extract corresponding embeddings for each modality.

For audio-head, we employ the ESResNeXt~\cite{guzhov2021esresne} model which has demonstrated a powerful ability to learn robust Time-Frequency transformation of audios. Firstly, the time-domain signal is converted to a log-power spectrogram using Short-Time Fourier Transform (STFT). Then, we feed the spectrogram into a 2-D convolutional layer with kernel size of $7 \times 7$ and a batch normalization layer. Next, 4 dual-path blocks~\cite{guzhov2021esresnet} are stacked to extract deep representation. In each dual-path block, the first path contains 3 stacked bottlenecks~\cite{he2016deep} with the structure shown in the yellow block. The second path consists of a max-pooling layer, 2 convolutional layers with kernel size of $3 \times 3$ and $1 \times 1$ respectively, and a batch normalization layer. The outputs of the two paths are multiplied as they have the same shape. Finally, the hidden features are fed into an average pooling layer and a linear layer to generate an audio embedding with a length of $L$. \par

For text-head, the caption input $t$ with any length is firstly converted into a word embedding. Then, a 12-layer Transformer~\cite{vaswani2017attention} is employed to extract the deep features of input embeddings. Next, a further normalization layer is added after the final attention block~\cite{radford2019language}. To obtain a text embedding with the same shape as the audio embedding, another linear layer is employed to map the features to embeddings with a length of $L$.

The parameters of the CLIP-AAC encoder are initialized by AudioCLIP~\cite{guzhov2021audioclip}, which includes three heads of image, audio and text. We remove the image-head and finetune the text-head and audio-head to adapt the AAC task. 
\begin{table*}[h]
\centering
\caption{BLEU{\scriptsize n}, METEOR, ROUGE{\scriptsize L}, CIDEr, SPICE, and SPIDEr in a comparative study of initialization strategies for the proposed CLIP-AAC encoder. All models are trained with batch size of 8.}
\resizebox{0.95\textwidth}{!}{
\begin{tabular}{c|c|c|c|c|c|c|c|c|c|c|c}
\hline\hline
System ID & Initialization & Frozen & BLEU{\scriptsize 1} & BLEU{\scriptsize 2} & BLEU{\scriptsize 3} & BLEU{\scriptsize 4} & ROUGE{\scriptsize L} & METERO & CIDEr & SPICE & SPIDEr \\ \hline\hline
1  & Random     & No &0.421 &0.246&0.159&0.070&0.312&0.103&0.060&0.059&0.060\\ \hline
2  & Audio-only & No & 0.480 &0.285 & 0.183 &0.114&0.326&0.132&0.217&0.081&0.149 \\ \hline
3  & Audio-Text  & Yes    & 0.521 &0.335 & 0.225 &0.147&0.356&0.153&0.333&0.100&0.217        \\ \hline
4  & Audio-Text  & No     & \textbf{0.561} & \textbf{0.365} & \textbf{0.245} &\textbf{0.161}&\textbf{0.372}&\textbf{0.168}&\textbf{0.394}&\textbf{0.115}& \textbf{0.254}         \\ \hline\hline
\end{tabular}}
\label{pretrain_model}
\end{table*}
\subsection{Contrastive learning}
We now introduce contrastive learning to learn the correspondence between audio embeddings and text embeddings. Specifically, the audio-head $Audio\_Enc$ and text-head $Text\_Enc$ are trained jointly 1) to maximize the cosine similarity of audio embeddings $A_i$ with corresponding text embeddings $T_i$ in a minibatch of $b$ pairs, and 2) to minimize the cosine similarity of the audio embeddings $A_i$ with negative examples. Such negative examples are remaining $b^2-b$ mismatched pairs. As shown in Figure ~\ref{CLIP-AAC}, we denote the similarities of matched pairs as purple cells that are distributed in the diagonal elements of the grid, while the mismatched pairs are labelled as grey cells in the grid. Such optimization is achieved by contrastive loss~\cite{zhang2020contrastive} over similarity scores, as shown in Algorithm~\ref{compute_audio_text_loss}.

\begin{algorithm}[h!]
  \caption{Pseudocode for Contrastive Loss}
  \begin{algorithmic}[1]
    \Require
      A minibatch of $b$ audio-text pairs $\{(a_i,t_i)\}$, $i \in \{1, 2, ..., b \}$
    \Ensure
      Contrastive loss $L_{cl}$
    \State Initialize the parameters of audio-head $Audio\_Enc$ and text-head $Text\_Enc$.
    \State Initialize temperature parameter $t$.
    \State Extract embeddings from each modality in minibatch:\par
        $A_i=Audio\_Enc(a_i), \ A_i \in \mathbb{R}^{1 \times L}$, \  \par
        $T_i=Text\_Enc(t_i), \ E_i \in \mathbb{R}^{1 \times L}$, \
         
    \State \textbf{for} i in 1, 2, ..., b \textbf{do} \par
      \textbf{for} k in 1, 2, ..., b \textbf{do} \par
    \qquad Calculate cosine similarity between $A_i$ and $T_k$\par
      \vspace{-0.1in}
    \begin{equation}
      \left\langle A_i,T_k \right\rangle=\frac{A_i \cdot (T_k)^T}{\left\|A_i\right\| \left \| T_k\right \|}
      \label{cs_audio}
    \end{equation} \par
        \textbf{end for}\par
    Calculate cross-entropy based contrastive loss for $A_i$
    \vspace{-0.1in}
    \begin{equation}
      L^{A\rightarrow T}_i = -\log \frac{exp(\left\langle A_i,T_i \right\rangle/t)}{\sum_{k=1}^b exp(\left\langle A_i,T_k \right\rangle/t)}
      \label{audio_loss}
    \vspace{-0.1in}  
    \end{equation} 
\textbf{end for}\par
    \State \textbf{for} i in 1, 2, ..., b \textbf{do} \par
      \textbf{for} k in 1, 2, ..., b \textbf{do} \par
      \qquad Calculate cosine similarity between $T_i$ and $A_k$\par
      \vspace{-0.1in}
    \begin{equation}
      \left\langle T_i,A_k \right\rangle=\frac{T_i \cdot (A_k)^T}{\left\|T_i\right\| \left \| A_k\right \|}
      \label{cs_text}
    \end{equation} \par
        \textbf{end for}\par
    Calculate cross-entropy based contrastive loss for $T_i$
    \vspace{-0.1in}
    \begin{equation}
      L^{T\rightarrow A}_i = -\log \frac{exp(\left\langle T_i,A_i \right\rangle/t)}{\sum_{k=1}^b exp(\left\langle T_i,A_k \right\rangle/t)}
      \label{text_loss}
    \vspace{-0.1in}
    \end{equation} 
\textbf{end for}\par
    \State Return the contrastive loss $L_{cl}$ \par 
    \vspace{-0.18in}
    \begin{equation}
      L_{cl} = \frac{1}{b} \sum^b_{i=1}(\lambda L^{A\rightarrow T}_i + (1-\lambda) L_i^{T\rightarrow A})
      \label{at_loss}
      \vspace{-0.1in}
    \end{equation}
  \end{algorithmic}
\label{compute_audio_text_loss}
\end{algorithm}

Given a minibatch of $b$ audio-text pairs, step 1 to 3 in Algorithm~\ref{compute_audio_text_loss} explains the feed forward process of audio and text inputs in the CLIP-AAC encoder. The $A_i$ and $T_i$ are the extracted embeddings from audio $a_i$ and its true caption $t_i$, respectively. In step 4, for each embedding $A_i$, we calculate its cosine similarities with every text embedding $\{T_1,T_2,...T_b\}$ in E.q (\ref{cs_audio}). Since only the true-pair $\left\langle A_i,T_i \right\rangle$ is viewed as positive example, we calculated the cross-entropy loss for each $A_i$ in E.q (\ref{audio_loss}). Similarly, a symmetric cross-entropy loss for each $T_i$ is calculated in E.q (\ref{cs_text}) and E.q (\ref{text_loss}). Finally, the contrastive loss $L_{cl}$ is computed by weight summing these two losses with a weight $\lambda$. We set $\lambda$ as 0.5 in this work. 

\subsection{Decoder}
The decoder of the proposed CLIP-AAC approach is a typical Transformer~\cite{vaswani2017attention} with multi-head attention, which has achieved state-of-the-art performance on various cross-modal tasks.\par
As shown in Figure~\ref{CLIP-AAC}, the tokens of the caption are firstly mapped to word embeddings and then fed into a masked multi-head self-attention layer to obtain hidden features. Subsequently, such hidden features are sent into another cross-attention layer to attentively fuse the audio embedding $A$ from the audio-head in the encoder. Considering the length of the caption, such an operation (grey block) is only repeated twice. Finally, a linear layer and softmax function are employed to produce the probability distribution along with the vocabulary.\par
During decoding, the $m$-th token $t_m$ is predicted based on previous tokens $\{t_0,t_1,...,t_{m-1}\}$ and audio embedding $A$, so the output probability of decoder is denoted as $P_\theta(t_m|t_0,t_1,...,t_{m-1}, A)$. The training objective is to maximize the log-likelihood for each predicted token via the cross entropy criterion $L_{ce}$:
\begin{equation}
  L_{ce} = -\sum_{m=0}^M log P_\theta(t_m|t_0,t_1,...,t_{m-1},A)
  \label{decoder_output}
\end{equation} 
Combined with contrastive loss $L_{cl}$ in E.q~\ref{at_loss}, we define the total loss $L_{total}$ for the whole neural network as:
\begin{equation}
  L_{total} = \alpha L_{cl} + (1-\alpha)L_{ce}
  \label{final_loss}
\end{equation} 
where $\alpha$ is the coefficient to trade-off each objective.

\section{Experiments}
\subsection{Database}
We conduct experiments on the Clotho dataset~\cite{drossos2020clotho}. The Clotho data is collected from the Freesound archive, and the duration of clips ranges from 15 to 30 seconds. The captions of Clotho are annotated by 5 different annotators so that each audio clip contains 5 captions for diversity. The length of captions ranges from eight to twenty words. \par

Following the previous works~\cite{koh2021automated}, we divided the Clotho dataset into a training set (4884 audio clips with 24420 captions), and a test set (1045 audio clips with 5225 captions). Each audio clip combines one of five captions as an audio-text training pair. At the inference stage, the average performances are calculated between predicted captions and all five captions.

\begin{table*}[h!]
\centering
\caption{BLEU{\scriptsize n}, METEOR, ROUGE{\scriptsize L}, CIDEr, SPICE, and SPIDEr in a comparative study of the proposed contrastive loss. ``$\alpha$" is the coefficient for contrastive loss, and ``B.S." denotes the batch size during training.}
\begin{tabular}{c|c|c|c|c|c|c|c|c|c|c|c}
\hline\hline
System ID & $\alpha$ & B.S. & BLEU{\scriptsize 1} & BLEU{\scriptsize 2} & BLEU{\scriptsize 3} & BLEU{\scriptsize 4} & ROUGE{\scriptsize L} & METERO & CIDEr & SPICE & SPIDEr \\ \hline\hline
5  & 0     & 8     &0.551 &0.358 & 0.241&0.158 &0.369  &0.162   & 0.368 & 0.109 &0.239  \\ \hline
6  & 0.1   & 8     &0.556 &0.359 &0.240 &0.156 &0.367 &0.165 &0.379&0.115 &0.247        \\ \hline
7  & 0.2   & 4     &0.556 &0.365 &0.246 &0.160& 0.371 &0.163 &0.372&0.109 &0.241\\ \hline
8  & 0.2   & 8     & 0.561&0.365 &0.245 &0.161 &0.372  &0.168  &0.394&0.115&0.254        \\ \hline
9  & 0.2   & 16    & \textbf{0.572}&\textbf{0.379}& \textbf{0.257}& \textbf{0.169}&  \textbf{0.379} & \textbf{0.171}& \textbf{0.407}& \textbf{0.118} & \textbf{0.263}       \\ \hline
10  & 0.3   & 8     & 0.561 &0.366 &0.248 &0.165 &0.374 &0.166 &0.380 &0.114 &0.247\\ \hline\hline
\end{tabular}
\label{alpha_bs}
\end{table*}

\begin{table*}[h!]
\centering
\caption{BLEU{\scriptsize n}, METEOR, ROUGE{\scriptsize L}, CIDEr, SPICE, and SPIDEr in a comparative study of CLIP-AAC and other competitive techniques.}
\begin{tabular}{c|c|c|c|c|c|c|c|c|c}
\hline\hline
Method & BLEU{\scriptsize 1} & BLEU{\scriptsize 2} & BLEU{\scriptsize 3} & BLEU{\scriptsize 4} & ROUGE{\scriptsize L} & METERO & CIDEr & SPICE & SPIDEr \\ \hline\hline
GRU Baseline~\cite{drossos2020clotho}   & 0.389 & 0.136 & 0.055 & 0.015 & 0.262 & 0.084 & 0.074 & 0.033 & 0.054 \\ \hline
PreCNN Transformer~\cite{chen2020audio} & 0.534 & 0.343 & 0.230 & 0.151 & 0.356 & 0.160 & 0.346 & 0.108 & 0.227 \\ \hline
CL4AC~\cite{liu2021cl4ac}               & 0.553 &0.349  & 0.226 & 0.143 & 0.374 & 0.168 & 0.368 & 0.115 & 0.242 \\ \hline
AT-CNN10~\cite{xu2021investigating}     & 0.556 &0.363  & 0.242 & 0.159 & 0.368 & 0.169 & 0.377 & 0.115 & 0.246 \\ \hline
TL + RLSSR~\cite{koh2021automated}      & 0.551 &0.369  & 0.252 & 0.168 & 0.373 & 0.165 & 0.380 & 0.111 & 0.246 \\ 
\hline\hline
CLIP-AAC (ours)                         & \textbf{0.572}&\textbf{0.379}& \textbf{0.257}& \textbf{0.169} &  \textbf{0.379} & \textbf{0.171} & \textbf{0.407}& \textbf{0.119} & \textbf{0.263}\\ \hline\hline
\end{tabular}
\label{benchmark}
\end{table*}

\subsection{Network Configuration}
At the training stage, the network was optimized by Adam~\cite{kingma2014adam}. To prevent the exploding gradients, a small learning rate of $1\times10^{-5}$ is applied to the pre-trained encoder and $1\times10^{-3}$ is applied to the decoder. A warm-up strategy is applied in the first 5 epochs that linearly increases the initial learning rate. The training epoch is set to 40 and the default batch size is 8. In addition, we use label smoothing~\cite{szegedy2016rethinking} with $\epsilon=0.1$ and dropout with a rate of 0.2 to mitigate the over-fitting problem. During inference, the beam search is applied with a beam size of 3.\par

\subsection{Metrics}
We report performances on six metrics, including BLEU{\scriptsize n}~\cite{papineni2002bleu}, METEOR~\cite{lin2004rouge}, ROUGE{\scriptsize L}~\cite{lavie2007meteor}, CIDEr~\cite{vedantam2015cider}, SPICE~\cite{anderson2016spice}, and SPIDEr~\cite{liu2017improved}.
\begin{itemize}
    \item BLEU{\scriptsize n} measures the correspondence between generated text and reference text by computing the precision of $n$-gram in the text.
    \item METEOR measures the harmonic mean of precision and recalls based on word-level matches between generated text and reference text.
    \item ROUGE{\scriptsize L} computes the F-measures based on the longest common sub-sequence.
    \item CIDEr both considers term frequency inverse document frequency weights of n-grams and the cosine similarity between the generated text and reference text.
    \item SPICE converts the captions to scene graphs and computes F-score based on tuple in them.
    \item SPIDEr is the linear combination of CIDEr and SPICE.
\end{itemize}
The first three metrics are proposed for machine translation systems but are also widely used to evaluate AAC systems in the previous works~\cite{liu2021cl4ac,mei2021encoder,xu2021investigating,koh2021automated}. The last three metrics are specifically used for captioning task~\cite{vinyals2015show,zhou2019grounded,you2016image}. For all metrics, higher score denotes better performance.

\section{Results}

\subsection{Effect of Initialization strategies for CLIP-AAC encoder}
\label{4_1}
We first analyze the performances of various initialization strategies for the encoder in the proposed CLIP-AAC approach. As shown in Table~\ref{pretrain_model}, system 1 and system 2 only utilize the audio-head in encoders, while system 3 and system 4 combine the proposed audio-head and text-head in encoders. Specifically, ``Random" means that the encoder in system 1 is initialized randomly. ``Audio-only" denotes that the encoder in system 2 is pre-trained on Audioset. ``Audio-Text" depicts that the encoders in system 3 and 4 are initialized with the modified AudioCLIP and then fine-tuned on Audioset. Compared with system 3, the encoder in system 4 is learnable during the training procedure.

We observe from the performances of system 1 that the encoder may not produce good features in deep architectures with random initialization. Compared with system 2, system 3 achieves better performance as the proposed encoder with audio-head and text-head can learn both acoustic and textual information. System 4 further improves performances by making the encoder trainable on the Clotho dataset. We adopt the setting of system 4 for the encoder hereafter.

\subsection{Effect of Contrastive Loss}
We further report the effect of proposed contrastive loss $L_{cl}$ on CLIP-AAC approach. As shown in Table~\ref{alpha_bs}, $\alpha$ is the coefficient to trade-off the contrastive loss $L_{cl}$ and cross-entropy loss $L_{ce}$. $B.S.$ denotes the batch size used at training stage. 

From system 5, 6, 8 and 10, we observe that performances improve as the coefficient $\alpha$ increases, indicating that the contrastive learning strategy can learn more audio and text information. We obtain best performances when $\alpha$ for $L_{cl}$ is set as 0.2. Furthermore, system 7-9 improve performances by increasing the batch size $B.S.$ because a large batch size can increase the diversity of negative audio-text pairs for training. Best performances are obtained with $B.S.$ of 16.

\begin{figure}[t]
\begin{center}
\includegraphics[scale=0.39]{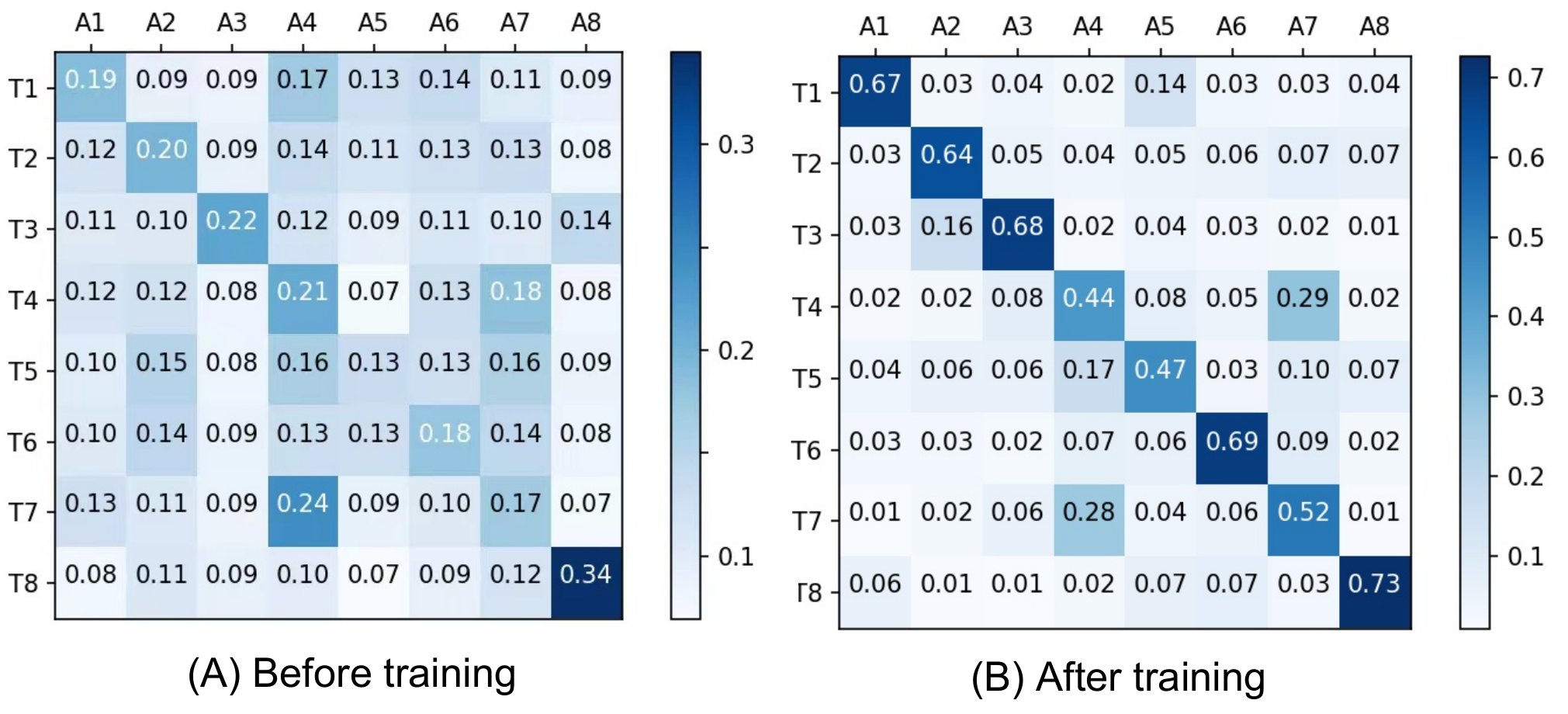}
\end{center}
\vspace{-0.15in}
\caption{Visualization of cosine similarity (after softmax) matrix between audio embeddings ($A_1$ to $A_8$) and text embeddings ($T_1$ to $T_8$). (A) and (B) denote the system 8 before and after training, respectively.}
\vspace{-0.2in}
\label{f1}
\end{figure}

To further show the contribution of the contrastive loss, we visualize the cosine similarity matrix of audio-text embeddings after the softmax function in system 8. The diagonal elements denote the cosine similarity of true audio-text pairs in Figure~\ref{f1}. We observe that the proposed pre-trained CLIP-AAC encoder can catch some correspondence between true audio-text pairs before training. After training, such correspondence is significantly enhanced by proposed contrastive learning. Therefore, we draw a conclusion that the CLIP-AAC encoder can provide high-quality embeddings for cross-modal decoding.

\subsection{Benchmark against other competitive methods}
Table~\ref{benchmark} summarizes the comparison between the proposed CLIP-AAC and other competitive techniques in terms of BLEU{\scriptsize n}, METEOR, ROUGE{\scriptsize L}, CIDEr, SPICE, and SPIDEr. We conduct experiments on the Clotho dataset. Except for the GRU baseline~\cite{drossos2020clotho}, all encoders in the other five techniques are pre-trained on AudioSet. The GRU baseline is trained from scratch. We observe that the proposed CLIP-AAC obtained the best performances for all metrics.

\section{Conclusions}
In this paper, we propose CLIP-AAC to learn interactive cross-modality representation with both acoustic and textual information for AAC tasks. Specifically, we introduce an audio-head and text-head in the encoder to capture audio-text correspondence. Furthermore, we also propose a contrastive learning strategy to optimize the CLIP framework. Experimental results show the proposed CLIP-AAC outperforms the best baseline by a large margin in terms of all metrics.  

\section{Acknowledgements}
This research is supported by the National Research Foundation, Singapore under its AI Singapore Programme (AISG Award No: AISG-100E-2018-006). The computational work for this article was partially performed on resources of the National Supercomputing Centre, Singapore (https://www.nscc.sg).
\newpage

\bibliographystyle{IEEEtran}

\bibliography{mybib}


\end{document}